\documentclass[preprint,showpacs,preprintnumbers,amsmath,amssymb,floats,groupedaddress]{revtex4-1} 
\usepackage{graphicx}
\usepackage[english]{babel}
\usepackage{amsmath}
\usepackage{amssymb}
\usepackage{color}
\usepackage{bm}

\newcommand{\be}{\begin{eqnarray}} 
\newcommand{\ee}{\end{eqnarray}}

\newcommand{\cg}{c_{{\bf k+g}}}
\newcommand{\cgd}{c^+_{{\bf k+g}}}

\begin{document}
\title{Pseudogap and Fermi-arcs in underdoped cuprates}

\author{Chandra M. Varma$^*$}
\affiliation{Department of Physics, New York University, New York, NY 10003, \\
University of California, Berkeley, CA. 94704}

\begin{abstract}
The proposed loop-current order in cuprates  cannot give the observed pseudogap and the Fermi-arcs because it preserves translation symmetry. A modification to a periodic arrangement of the four possible orientations of the order parameter with a large period of between about 12 to 30 lattice constants is proposed and shown in a simple and controlled calculation to give one-particle spectra with every feature as in the ARPES experiments. The results follow from (1) the currents at the boundaries of the periodic domains with similar topology as the Affleck-Marston flux phase, and (2) the mixing introduced by the boundary currents  between the states near the erstwhile Fermi-surface and the ghost Fermi-surfaces which are displaced from it by mini-reciprocal vectors. The proposed idea can be ruled out or verified by high resolution diffraction or imaging experiments. 
It does not run afoul of the variety of different experiments consistent with the loop-current order as well as the theory of the marginal Fermi-liquid and d-wave superconductivity based on quantum-critical fluctuations of the loop current order.\end{abstract}
\date\today
\maketitle
*Visiting Scientist

\section{Introduction}

The cuprate phase diagram \cite{Keimer-rev2015} has presented three new phenomena in physics, two new normal states - the strange metal and the pseudogap with Fermi-arcs \cite{ARPES-revZX, Campu-rev}, and the high temperature d-wave superconducting phase. It is now generally accepted that the pseudogap occurs below a phase transition at $T^*(x)$ ending at a quantum-critical point as a function of doping $x = x_c$, as was predicted \cite{cmv1997}. The strange metal region occurs on the other side of $T^*(x)$ and ends in a gradual cross-over to a Fermi-liquid phase. The Fermi-liquid normal phase further supports the existence of
a quantum-critical point. Superconductivity occurs in a region of $x$ around the quantum-critical point.

This paper is concerned with the unanswered questions about the one-particle spectra in the pseudogap state in the cuprate metals. An angle-dependent gap is observed in ARPES experiments \cite{ARPES-revZX, Campu-rev} at the erstwhile Fermi-surface (measured above $T^*(x)$) decreasing in magnitude from the $(0,\pi)$ directions ending in a Fermi-arc - a region with zero gap over a finite angle centered in the $(\pi,\pi)$ directions. In the same region of the phase diagram, a tiny closed Fermi-surface, about 2\% of that expected by calculated band-structure, is observed in magneto-oscillation experiments at low temperatures and high enough magnetic fields \cite{Proust-Sebastian2015}. Refined experiments \cite{Sebastian2012} suggest that such small Fermi-surfaces occur near the diagonal directions where ARPES observes Fermi-arcs. 

$T^*(x)$ does mark the onset of the loop-current ordered state \cite{simon-cmv} which breaks time-reversal symmetries as well as reflection and rotation symmetries. Seven different experimental techniques \cite{Bourges-rev, Kaminski-diARPES, Leridon, Hsieh2017, Kapitulnik1, Armitage-Biref,  muons1210, Shekhter2013, Matsuda-torque1, Matsuda-torque2} which test different aspects of such symmetries, are consistent with the occurrence of such a state below $T^*(x)$.  Quantum critical fluctuations \cite{ASV2010} of this state coupled to fermions  give in a systematic theory the marginal fermi-liquid \cite{CMV-MFL} universally observed in a variety of experiments in the strange metal phase. The spectra of these quantum fluctuations is deduced through analysis of high resolution ARPES measurements  \cite{Bok_ScienceADV} to  to give d-wave superconductivity as well as the one-particle spectra  of the marginal Fermi-liquid. These lead to essentially all the anomalies observed in the strange metal region including those in the particle-hole channels. For example, the observed density fluctuations in the cuprates, radically different from ordinary metals at almost all momenta follow \cite{Abbamonte2018} as well as the Raman response \cite{Hackl-rev} in various irreducible representations. 

But this state does not change the translational symmetry and so it cannot lead to the observed electronic structure below $T^*(x)$. However since the pseudogap appears below the same temperature $T^*(x)$ at which the various experiments show a broken symmetry consistent with loop-current order \cite{simon-cmv, Weber-Mila, Weber-Giam-V}, and since it has a large condensation energy, estimated \cite{CMV-Zhu-PNAS} from the magnitude of the order parameter of about $0.1 \mu_B$ per unit-cell found by polarized neutron scattering to be similar to that deduced in experiments for the pseudogap state, it is quite unlikely that entirely different physics is required to explain it.

The essential idea used in this work is that the loop-current order parameter has the symmetry of a vector potential, or more appropriately of the gauge invariant quantity - (intra-unit-cell) current. A periodic variation of the four different orientations of the order parameter forming a super-cell is equivalent to loop currents at the domain boundaries. The periodic problem is then a variant on the Hofstadter problem \cite{Hofstadter}. The periodic variation is in fact favored energetically. It is shown by a simple controlled calculation using two small parameters, a parameter  $\phi <<1$ related to the phase difference across the nearest neighbor bonds and the inverse period $1/2P <<1$, that such a periodic loop-current order leads to a one-particle spectra consistent with the symmetry and magnitude of the pseudogap and the Fermi-arcs observed by ARPES. 

The new idea is consistent with the experiments which have observed loop-current order without apparent alteration of translation symmetry within their present resolution. It also does not introduce features which are ruled out by any available experiments.
The present  neutron scattering experiments  put a lower limit to the periodicity of about 8 $\times$ 8 lattice constants. Experiments to look for longer periods with very high resolution diffraction are suggested. The applicability of the present idea and calculations lives or dies depending on the outcome of such experiments.

An important aspect to the pseudogap is to be learnt from the succesful fit to the ARPES data provided very early by Norman, Campuzano, Ding and Randeria \cite{Norman}. Motivated by the idea that the pseudogap may be due to preformed BCS pairs, a  d-wave BCS spectral function with a large relaxation rate was proposed. 
Various details of such an ansatz have been tested in even more detail recently \cite{Kaminski-pg}.
We know now for a number of years from direct experiments that there are not even noticeable superconducting fluctuations for temperatures below $T^*(x)$ to about 20 K above $T_c(x)$.
But the success of the fit provides the lesson that one must have an angle-dependent gap tied to the entire erstwhile Fermi-surface as in the BCS theory. The BCS theory of-course is based on the perfect nesting in the particle-particle channel of states ${\bf p}$ and ${\bf -p}$ near the Fermi-wave-vectors. (The spin-labels will be dropped where unnecessary.) No other phase is known fulfilling this condition. Other well known phases, charge density waves or spin-density waves have gaps related to the magnitude and symmetries of the nesting wave-vectors ${\bf Q}$ besides the Fermi-surface. In a two-dimensional fermion problem, they give a gap all around the Fermi-surface (except at $(\pm (\pi/2,\pi/2))$ only for nearest neighbor hopping on a square lattice at half-filling. In general, they produce many different gaps and/or open Fermi-surfaces which are not observed in the cuprates either in ARPES or magneto-oscillations \cite{Proust-Sebastian2015}. Moreover no CDW or SDW with large enough correlation length and amplitude to give the magnitude of the gap observed is known to exist universally in the pseudogap region. Small Fermi-surfaces do arise in models of CDWs with finite fields, likely giving the small Fermi-surfaces observed in magneto-oscillation experiments \cite{Sebastian2012}. But an independent unknown mechanism must then be invoked \cite{Proust-Sebastian2015} for the lack of observation of the additional Fermi-surfaces, closed or open which are inevitably predicted.

Suppose the period of the envisaged variation of the loop-current order is $2P \times 2P$, and $P >> 1$. 
The gap due to the periodic variations of Loop-current order will be shown to be tied to the erstwhile Fermi-surface ${\bf p}_F$ to the accuracy better than $(1/2P){\bf p}_F$. The geometry of the currents at the domain boundaries is such that the gap also has the angular dependence of the observed pseudogap which vanishes in the clean limit vanishes in the $(\pi,\pi)$ directions and has an angular width (the Fermi-arc) related to the linewidth. 
Additional features, $O(P^2)$ in number, besides the principal feature with the pseudo-gap and the Fermi-arc on which we concentrate in this paper, occur but the spectral weight in any of them on the average is proportional  to $\phi / P^2$. These aspects are all shown and explained by calculations below.

\section{Periodic Variation of Loop-Current Order}
\begin{figure} 
 \includegraphics[width=1.0\textwidth]{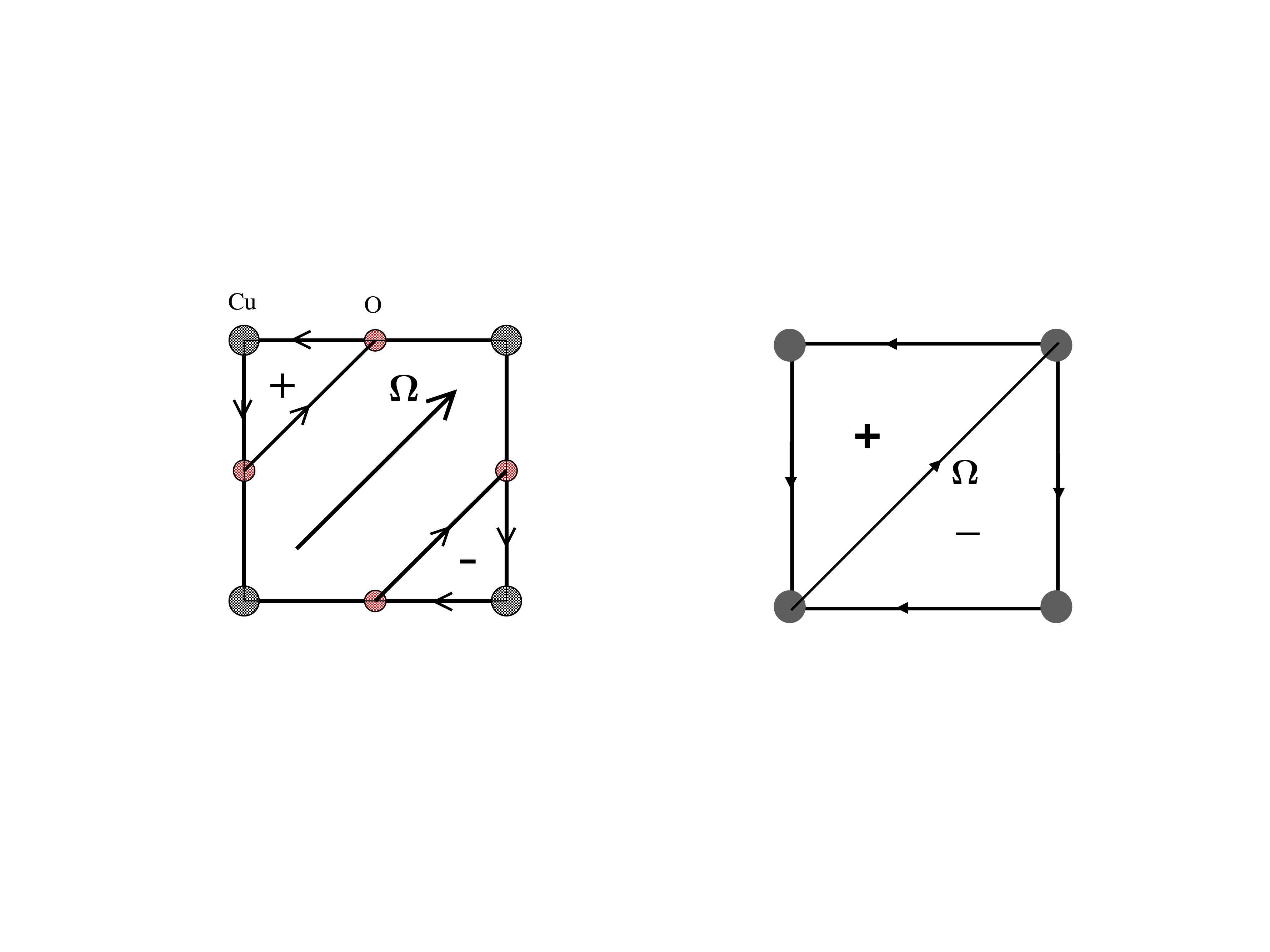}
\caption{Left: Representation of Loop-current order in Cuprates. Right: The simplified current pattern of Loop-current order with the same symmetries as on left.}
\label{Fig1}
\end{figure}

 The vector potential (or current in bonds) representing the previously proposed loop-current order \cite{simon-cmv}  is sketched in Fig. (\ref{Fig1}-left). In a unit-cell this has the same symmetry as the simpler representation \cite{Berg2008} in Fig. (\ref{Fig1}-right), which is adopted for the calculations. The order parameter ${\bf \Omega}$, shown as an arrow in the figure is odd under time-reversal and under inversion and so has the symmetry of a current or a vector potential. The flux integrated over a unit-cell is zero. 
 
\begin{figure}
\includegraphics[width=0.75\textwidth]{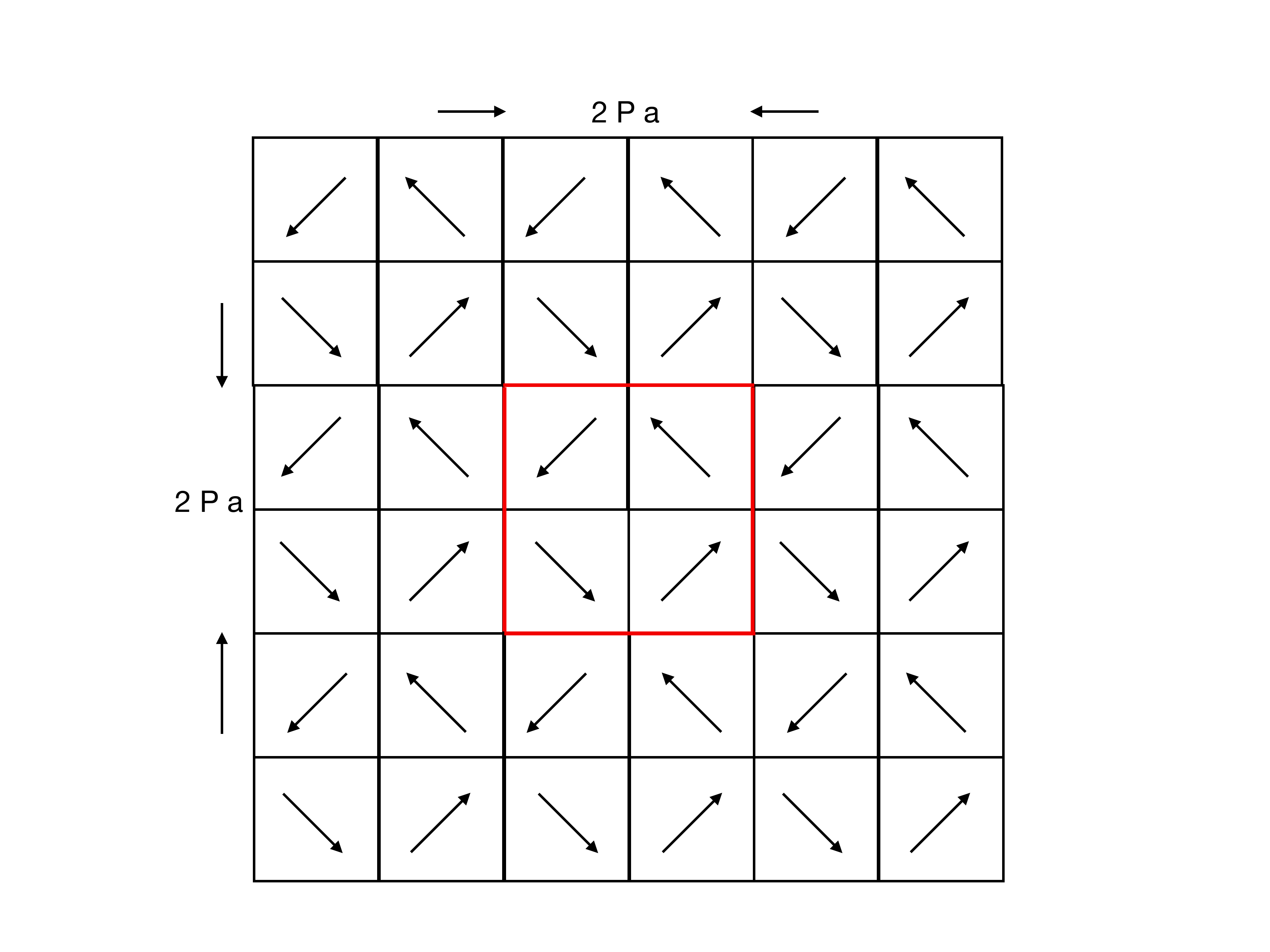}
\caption{The arrangement of the four domains as a periodic lattice. The red square indicates the super unit cell of size
$2Pa \times 2Pa$. The intersection of the four domains form a periodic arrangement of topological defects of vector field ${\bf \Omega}$.}
\label{Fig2}
\end{figure}

There are four directions in which ${\bf \Omega}$ can lie. A modification of the translation preserving order is considered whereby the four possible domains of order are arranged in a periodic pattern of period $2 Pa$ in both $x$ and $y$ directions, as in Fig. (\ref{Fig2}). 

\subsection{Hamiltonian}

The loop-current order was shown to be a possible ground state of the three orbital model for cuprates \cite{simon-cmv}  by expressing the electron interaction operator between sites in terms of product of current operator on bonds. A mean-field theory with the expectation value of the current operators organized in specific symmetries on the bonds as the order parameter was shown to have a stationary point. Such a calculation is not repeated here. We will assume the same many-body Hamiltonian and assume that a local minima with translational symmetry preserved arises as a starting point. We then
consider the re-organization of the four possible domains of the loop-current order in a periodic manner with period $2P \times 2P$ and calculate the one-particle spectra. We assume the magnitude of the current in the bonds is similar to what was obtained earlier or deduced from the magnitude of the moment discovered by polarized neutron scattering. 

Let us represent the lattice points by $(i_x,n_x; i_y,n_y)$. $(i_x,i_y)$ denote the points on the super-cell and in each super-cell $(n_x, n_y)= (-P,..,P)a$ denote the lattice points of the original unit-cells in the super-cell. The Hamiltonian has two parts $H_0$ and $H_1$. We take $H_0$ just to be the one-particle kinetic energy for fermions on the square lattice. $H_1$ introduces the domains of loop-current order.
\be
\label{Ham}
H &=& H_0+H_1 \\
H_0 &=&  \frac{1}{N^2}\sum_{(i_x,n_x, i_y,n_y)}-t \big( c^+_{(i_x,n_x, i_y,n_y)}c_{(i_x,n_x+1, i_y,n_y)} + c^+_{(i_x,n_x, i_y,n_y)}c_{(i_x,n_x, i_y,n_y+1)} \big) \\ \nonumber
&+& t' ~\big( c^+_{(i_x,n_x, i_y,n_y)}c_{(i_x,n_{x+1}, i_y,n_{y+1})}  + c^+_{(i_x,n_{x+1}, i_y,n_y)}c_{(i_x,n_{x}, i_y,n_{y+1})}\big) + H.C.
\ee
 $t$ is the coupling of the nearest neighbor horizontal and vertical nearest neighbor bonds and $t'$ is the coupling of the diagonal next-nearest neighbor bonds. $N^2$ is the number of unit-cells. $t' \approx 0.6 t$ correctly gives the Fermi-surface observed above $T^*(x)$ in Bi2212 with about 12\% deviation from 1/2 filling.  A more general $H_0$ does not change the essential results. 
 
 $H_1$ is the Hamiltonian which introduces the four domains of the loop-current order. Each domain is on a $P \times P$ lattice and is arranged as in Fig. (\ref{Fig2}). 
 The currents in the bonds are given by $H_1$ obtained by adding an imaginary parts to the transfer integral $ i t \phi$, oriented appropriately. It is assumed, as is realistic from the estimate of the magnitudes of magnetic moment discovered by polarized neutron scattering, that $\phi/\pi <<1$. This is similar to the mean-field and better \cite{simon-cmv, Weber-Mila, Weber-Giam-V} calculations done earlier in which the kinetic contains a term proportional to the expectation value of the order parameter and a phase difference which is self-consistently determined to give the order parameter. So the final value of $\phi$ is proportional to the square of the order parameter. I have adopted a gauge in which the diagonal bonds do not carry a phase-factor.
As long as there are finite diagonal hopping terms, $t'$ in Eq. (\ref{Ham}) there is a clockwise flux and a counterclockwise flux each with magnitude proportional to $2\phi$ in two triangles in each unit-cell. The zero flux condition in each unit-cell allows periodic arrangement of the $\phi's$ or of the ${\bf \Omega}$'s.\\

\begin{figure}
\begin{center}
\includegraphics[width=1.0\textwidth]{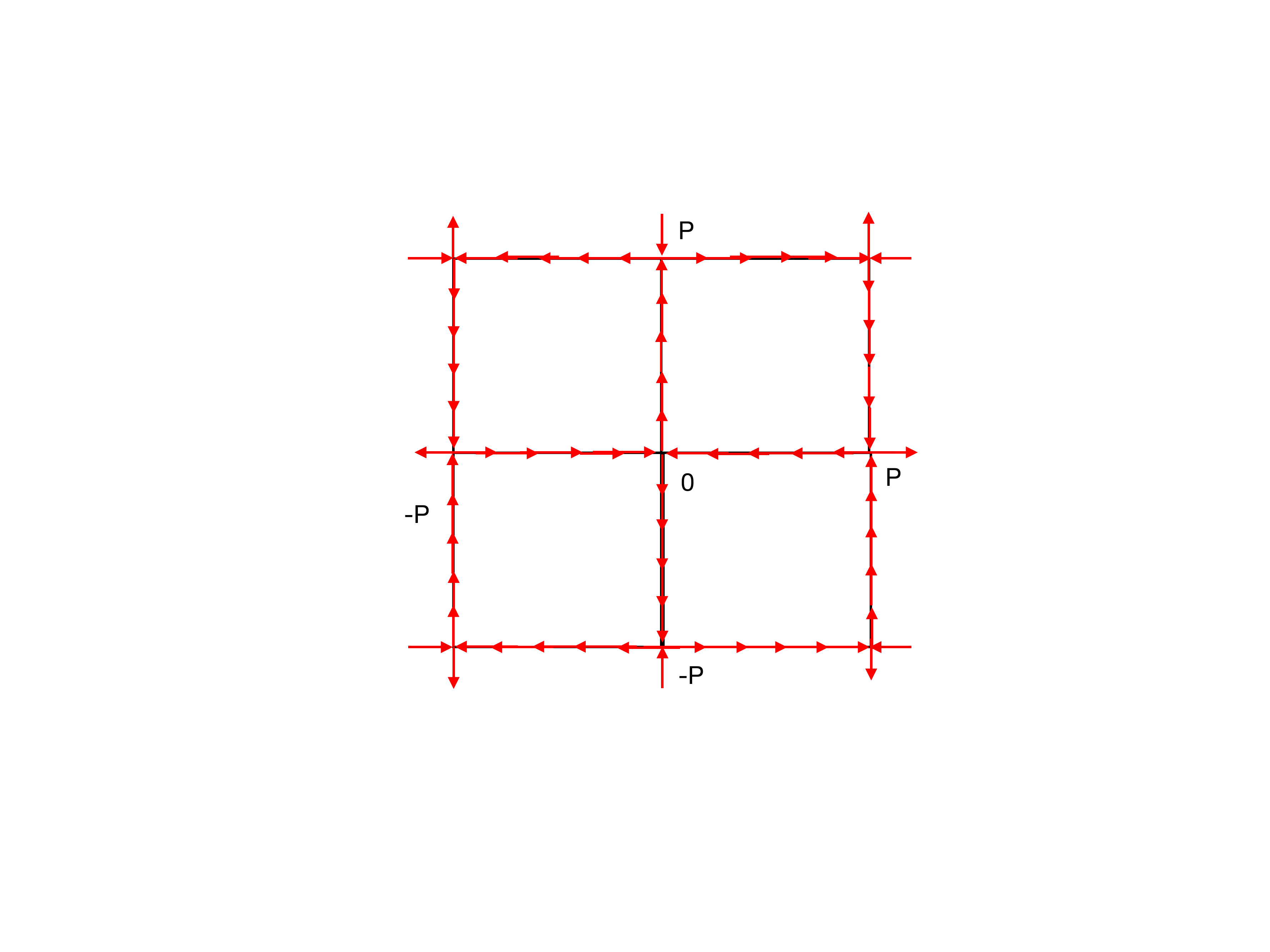}
\end{center}
\caption{Currents at Boundaries of domains oriented so that the direction of currents preserves the boundary current of each of the domains optimally. The figure illustrates the case of $P=5$ or $10 \times 10$ unit-cells in a super-cell.}
\label{Fig3}
\end{figure}

The four domains are characterized by the imaginary transfer integrals $\pm i \phi$ in the $x$ and $y$ bonds respectively as noted below in (\ref{ABCD}). Adjacent super-cells are share their common boundaries. Within each super-cell the four domains are arranged as follows:
\be
\label{ABCD}
A  ~&& (i\phi, i\phi) ~: (0 \leqslant n_x \leqslant P, ~ 0 \leqslant n_y \leqslant P);  \nonumber  \\
B  ~&& (-i\phi, i\phi) ~: -P \leqslant n_x \leqslant 0, ~ 0 \leqslant n_y \leqslant  P); \nonumber  \\
C  ~&& (-i\phi, -i\phi) ~: (-P \leqslant n_x \leqslant 0, ~ -P \leqslant n_y \leqslant P);  \nonumber  \\
D  ~ && (i\phi, -i\phi) ~ : (0 \leqslant n_x \leqslant P, ~ -P \leqslant n_y \leqslant  0). 
\ee
As written the bonds at $n_x=0$ for all $n_y$ are included in both domains A and B, and in C and D, and $n_y=0$  for all $n_x$ are included in both domains A and D, and in domains B and C. The intersection point $n_x = n_y =0$ is included in all four domains. Similar features occurs at the edges and corners of the super-cell. With this choice, there is no current along the domain boundaries; there is also no current across the domain boundaries because currents loops close within any unit-cell as in Fig. (\ref{Fig1}). It is useful to note that if there is no current at the domain boundaries, there are no closed loops of currents in the cells adjoining them. In that case, through a gauge transformation, currents can be eliminated in the boundaries of those cells and successively in the entire region. In effect, since there is zero flux integrated over any unit-cell, eliminating the phase differences (or vector-potential) at the boundaries in effect eliminates them everywhere. In the continuum limit the coarse grained vector potential in the present case is the gradient of a scalar. So eliminating it at the boundary enclosing a region allows shrinking the boundary as no singularities are found (as long as only distances of a unit-cell or larger are considered). If this is correct, there can be no scattering among states of $H_0$ due to $H_1$ with the choice of boundary currents made above. This will be shown explicitly. This implies that the symmetry of the four different ${\Omega}$ in the domains and their periodic variation must be introduced though proper choice of phase differences or currents at the boundaries.  The currents in the domain boundaries are chosen as shown in Fig. (\ref{Fig3}). This choice cyclically preserves the ${\bf \Omega}$ in the cells adjoining two of the boundaries in each domain.

$H_1$ in regions of space occupied by domain A is given by,
\be
\label{H1-A}
H_1(A) = i t \phi ~\frac{P^2}{N} ~\sum_{i_x,i_y} \frac{1}{4P^2}  \sum'_{(n_x = 0,..,P; n_y = 0,..,P)} & & c^+_{(i_x,n_x, i_y,n_y)} \nonumber  \\
& & \times \big( c_{(i_x,n_{x+1}, i_y,n_y)} +  c_{(i_x,n_x, i_y,n_{y+1})} \big) +H.C.
\ee
The restriction in the sum in $H_1(A)$ specifies $\phi$ in the bonds at the boundaries. The bonds in the $y-$ direction are removed from the sum in (\ref{H1-A}) at $n_x=0$ for $0 \leqslant n_y \leqslant P$, and in the $x-$ direction are removed at $n_y=0$ for $0 \leqslant n_x \leqslant P$. The other two boundaries are left un-changed. Together with the restrictions on the adjoining domains, the currents along the boundaries are then as shown in Fig. (\ref{Fig3}). 
\be
\label{H1-B}
H_1(B) = i t \phi  \frac{P^2}{N} \sum_{i_x,i_y} \frac{1}{4P^2} \sum'_{(n_x = -P,..,0; n_y =0,.., P)} & & c^+_{(i_x,n_x, i_y,n_y)} \nonumber  \\
& & \times \big(- c_{(i_x,n_{x+1}, i_y,n_y)} + c_{(i_x,n_x, i_y,n_{y+1})} \big) + H.C.
\ee
Restrictions on the sum in $H_1(B)$, for the other two domains are devised as for $H_1(A)$ to maintain the current at the boundaries as in Fig. (\ref{Fig3}).
\be
\label{H1-C}
H_1(C) = i t \phi  \frac{P^2}{N} \sum_{i_x,i_y} \frac{1}{4P^2}\sum'_{(n_x = - P,..,0; n_y = -P,.., 0)} & & c^+_{(i_x,n_x, i_y,n_y)} \nonumber \\
& & \times \big(- c_{(i_x,n_{x+1}, i_y,n_y)} - c_{(i_x,n_x, i_y,n_{y+1})} \big) + H.C.   \nonumber \\
\label{H1-D}
H_1(D) = i t \phi  \frac{P^2}{N} \sum_{i_x,i_y} \frac{1}{4P^2}\sum'_{(n_x = 0,..,P; n_y = - P,..,0)} & & c^+_{(i_x,n_x, i_y,n_y)}  \nonumber  \\
& & \times \big(c_{(i_x,n_{x+1}, i_y,n_y)} - c_{(i_x,n_x, i_y,n_{y+1})} \big) + H.C.
\ee

\subsection{Hamiltonian in momentum basis}

Now we transform to momentum space. The Bravais  lattice is of size $2 Pa \times 2 Pa$, (and not $\sqrt{2}  P a \times \sqrt{2}  P a)$, and the mini-Brillouin zones (mBZ) are correspondingly $1/(2P)^2$ times smaller. To describe the entire momentum space, one may choose the quantum-numbers to be ${\bf k, g} \equiv (k_x,k_y, g_x, g_y)$.  ${\bf g} \equiv \Big(\frac{2 \pi \nu_x}{2 P a}, \frac{2 \pi \nu_y}{2 P a}\Big); (\nu_x = 1,...P, \nu_y = 1,..P)$ are the reciprocal vectors which are
used to extend the vectors into the previous big BZ. A vector ${\bf p}$ in the original  big BZ, with $-2\pi/(Pa) \leqslant (p_{x}, p_{y}) \leqslant 2\pi/(2a)$ is described by 
\be
&{\bf p} \equiv  {\bf g}_{\nu_x, \nu_y} + {\bf k}, \\ \nonumber
-\frac{2\pi}{2Pa}&(\nu_x,\nu_y) \leqslant (k_{x}, k_{y}) \leqslant \frac{2\pi}{2Pa}(\nu_x,\nu_y).
\ee
So ${\bf g}_{\nu_x, \nu_y}$ is the center of a mBZ $(\nu_x, \nu_y)$. One should add a sub-script $(\nu_x,\nu_y)$ to ${\bf k}$ to indicate which mBZ it belongs. But in order to avoid too much clutter I do not do that.

  $H_0$ diagonalizes to 
\be
\label{H0}
H_0 = \sum_{{\bf k,g}} \epsilon_{0}({\bf k,g}) \cgd \cg.
\ee
$\epsilon_0$ is periodic in ${\bf g}$ for any ${\bf k}$. For any vector ${\bf p}$, there is a privileged or principal mBZ ${\bf g}_p$.
 It is privileged in the sense that for $H_1 =0$ the weight $<{\bf k +g_p}|{\bf k +g_p}> = 1$, while 
similarly defined weight is $0$ for $ {\bf g} \ne {\bf g}_p$. These simple facts are being stated because their use is essential in later developments.

Now, we consider the scattering from ${\bf k, g}$ to ${\bf k, g'}$ due to the domains formed due to $H_1$. First it is shown that this scattering occurs only due to the vector potentials introduced at the domain boundaries.
This is done by first ignoring the boundary restrictions on $H_1$ and show that without the restrictions, it has no contribution to scattering.
In expressing $H_1$ in momentum space, the following identities are used.
\be
\frac{(P)^2}{N}\sum_{{\bf R}(i_x,i_y)} e^{i ({\bf k -k'}).{\bf R}} &=&\delta({\bf k -k'}); \\
e^{i {\bf g}\cdot {\bf R}(i_x,i_y)} &=& 1.
\ee
${\bf R}(i_x,i_y)$ is any lattice vector of the super cell. 

$H_1$ is necessarily off-diagonal in the ${\bf g}$. Let us first ignore the restrictions at the boundaries in all four $H_1$'s. Define
\be
F_{xx'}(P) \equiv \frac{1}{P} \sum_{n_x =0,...P} e^{i n_x (g_x-g'_x) a} &=&  e^{-i (\nu_x-\nu_{x'})\pi}~ \frac{1}{P} \sum_{n_x =-P,...,0} e^{i n_x (g_x-g'_x)a}.
 \ee
Similar relations hold for $F_{yy'}(P)$ obtained from  $F_{xx'}(P)$ by $(x \to y)$. $g_{x,y} \equiv (\nu_x, \nu_y) \frac{2\pi}{2Pa}$ and $(\nu_x, \nu_y) = 1,... 2P$.
The factor $e^{-i (\nu_x-\nu_{x'})\pi}$ is $(-1)^{|\nu_x-\nu_{x'}|}$  and so is $\pm 1$ for $|\nu_x-\nu_{x'}|$ even/odd. Also,
\be
\label{evenodd}
F_{xx'}(P) = \frac{1}{2P} \frac{e^{i (\nu_x-\nu_{x'}) \pi} -1}{e^{i (\nu_x-\nu_{x'})\pi/P} -1}  &= & 0, ~ for ~ |\nu_x-\nu_{x'}| ~ even,  \nonumber \\
   &= & -\frac{1}{P} \frac{2}{e^{i (\nu_x-\nu_{x'})\pi/P} -1}, ~ for ~ |\nu_x-\nu_{x'}| ~ odd. 
\ee
For later usage, it is worth noting that for small $(\nu_x-\nu_{x'})/P$, $F_{xx'}(P) = \frac{i}{2(\nu_x-\nu_{x'})}$ with corrections of $O((\nu_x-\nu_{x'})/P)$.

Using these relations, we first consider ${\overline{H}}_1= \overline{H}_1(A) + \overline{H}_1(B)+\overline{H}_1(C)+\overline{H}_1(D)$ which are the part of $H_1$ without the restrictions at the boundaries between the domains. In this case, as discussed, there is zero currents along all the boundaries.  ${\overline{H}}_1$ in momentum space is
\be
\label{H1-f}
{\overline{H}}_1 = i t ~ \phi \sum_{{\bf k}, {\bf g}, {\bf g}'} \cgd c_{{\bf k+g}'}  & F_{xx'}(P)F_{yy'}(P) \Big( e^{i(k_x +g'_{x})a} \big((1 - (-1)^{|\nu_x-\nu_{x'}|})(1+ (-1)^{|\nu_y-\nu_{y'}|})  \nonumber \\
&+  e^{i(k_y +g'_{y})a} \big((1 + (-1)^{|\nu_x-\nu_{x'}|})(1- (-1)^{|\nu_y-\nu_{y'}|})\big) \Big) + H.C.
\ee
The four signs in each of the terms come from the sign in front of $i\phi$ in $x$ and $y$ directions in each of the four domains. We now notice that that (\ref{H1-f}) gives zero because of the requirement in the parenthesis that   $|\nu_x-\nu_{x'}|$ or $|\nu_y-\nu_{y'}|$ be odd while the other is even (which is obvious on looking at Fig. (\ref{Fig2})). But the factors of $F$ in front are 0 for either  $|\nu_x-\nu_{x'}|$ or $|\nu_y-\nu_{y'}|$ even. 

So the only contribution to scattering from ${\bf k+g}$ to ${\bf k+g'}$ is through the boundaries.  Let us first write down the boundary Hamiltonian $H_b$ in real space, which is just the set of restrictions on $H_1$ before expressing it in $({\bf k, g})$ space.
There are 8 segments of boundaries per super-cell, four internal to the super-cell and four external. The restrictions impose oriented currents along the boundaries as in Fig. (\ref{Fig3}).
It is convenient before transforming to $({\bf k, g})$-space to arrange the boundary current Hamiltonian as follows:
\be
H_b = \frac{i t \phi}{P^2} & &\big(\sum_{n_x =0,..,P} c^+_{n_x,0}  c_{n_{x+1},0} +  \sum_{n_x =0,..,-P} c^+_{n_x,0}  c_{n_{x-1},0} -  \sum_{n_y =0,..,P} c^+_{0,n_y}  c_{0,n_{y+1},} -  \sum_{n_y =0,..,-P}c^+_{0,n_y}  c_{0,n_{y-1}}  \nonumber \\
-  & &  (n_y = P ~ {\textrm{ in first two terms above}}) -   (n_x=P  {\textrm{ in second two terms above}}) \nonumber  \\
\ee

We can now transform to ${\bf k, g}$ space. 
\be
\label{bH}
H_{b} = 2 \frac{i t \phi}{P} \sum_{{\bf k,g,g'}}  & &\big(F_{xx'}(P) (1-(-1)^{|\nu_y-\nu_{y'}|}) + F_{yy'}(P) (1-(-1)^{|\nu_x-\nu_{x'}|})\big) \nonumber \\ 
&&\big((\cos(k_x +g'_{x})a -  \cos(k_y +g'_{y})a)+ (\cos(k_x +g_{x})a-  \cos(k_y +g_{y})a )\big) \nonumber \\ &&
   \cgd c_{{\bf k+g}'} + H.C.
\ee
The relations $F^*_{xx'}(P) = F_{x'x}(P)$, etc., have been used to write $H_b$ in symmetric form between ${\bf g}$ and ${\bf g}'$.

The symmetry of the matrix elements of $H_b$ will be important. These symmetries are visible in Fig. (\ref{Fig3}). Both $(\nu_x-\nu_{x}')$ and $(\nu_y-\nu_{y}')$ must be odd for the matrix elements to be non-zero. From Eq. (\ref{evenodd}), the matrix elements decrease geometrically with increasing $(\nu_x-\nu_{x}')/P$ and $(\nu_y-\nu_{y}')/P$. The matrix element is $0$ if both $(k_x + g_{x}) = (k_y+g_{y})$ and  $(k_x + g'_{x}) = (k_y+g'_{y})$.
 
The arrangement of currents in Fig. (\ref{Fig3}) and Eq. (\ref{bH}) reminds one of the Affleck-Marston staggered flux in alternate unit-cells \cite{Affleck-Marston} of a square lattice which produces a scattering proportional to $(\cos(p_xa) -\cos(p_ya))$ between states of the two sub-lattices. The result in that case is a mass-less Dirac-cone spectrum of fermions at half-filling on a model with nearest neighbor kinetic energy alone, but it gives pieces of Fermi-surface both in the nodal and the anti-nodal directions away from half-filling or with next neighbor kinetic energy at all fillings. Echoes of Affleck-Marston are to be found in RVB with fluctuations of gauge fields put in \cite{Lee-Nagaosa-rev} and in the d-density wave order parameter \cite{Chakra-ddw}. The same symmetries and the same problems should occur with generation of Fermi-surfaces through alternating antiferromagnetic order or antiferromagnetic order with gauge-field fluctuations \cite{Georges-pseudogap, Tremblay, Georges-Sachdev}.  The multitudinous Fermi-surfaces generated with charge density waves with various periods have been well documented \cite{Proust-Sebastian2015}.  How Eq. (\ref{bH}), which gives small staggered flux in large super-cells does not have these problems will be evident below.

\section{One-particle spectral function}

We wish to derive the eigenvalues and the eigen-vectors of $H_0 + H_b$ and use them to calculate the one-particle spectral function. The exact solution entails diagonalizing a $(2P)^2 \times (2P)^2$ matrix in ${\bf g, g}'$ space for any ${\bf k}$. This is a formidable numerical task. But, the essentials of the problem can be reduced in a controlled way to diagonalizing a $5 \times 5$  matrix for any ${\bf p}$ by noting the following:\\
The Brillouin zone of $H_0$ has $2P \times 2P$ mBZ's. Among these mBZ's, there is a zone ${\bf g_p}$ which we have earlier called the principal mBZ. $({\bf k + g_p}) \equiv {\bf p}$.  For any eigenvalue $\epsilon_{0}({\bf k + g}_p)$ of $H_0$, the spectral weight of states in ${\bf g_p}$ is $1$ while states in the other mBZ's, ${\bf g}' \ne {\bf g_p}$ (in the extended zone) at the same energy have spectral weight $0$. So $|{\bf k+ g'}>$ acquires amplitude only due to $H_b$. Their amplitude in the eigenvectors of $H_0+H_b$ can therefore  be only of $O(1)$ if  $\epsilon_0({\bf k +g'})$ is nearly degenerate with $\epsilon_0({\bf k +g_p})$.  For energy difference large compared to the scattering matrix elements, their contribution to the spectral function is of $O(\phi^2)$. We will check below that there is no degeneracy for states between which there is scattering. A number of simplifications then follow. One needs only to calculate the scattering only between states $|{\bf k+g_p}>$ and $|{\bf k + g'}>, ~{\bf g'} \ne {\bf g}_p$, and not between the different such $|{\bf k+g'}>$. As shown above the matrix elements go down as $1/(P|\nu_x - \nu_x'|)$ for large $P$. We need therefore consider only scattering between $|\bf{k+ g}_p>$ and $|\bf{k+ g}'>$, such that $|\nu_{x,y} -\nu_{x',y'}| = 1$.   The next states with $|\nu_{x,y} -\nu_{x',y'}| =3$ and $P=9$ introduce corrections to eigenvalues which are about (1/27) or more smaller. 

\begin{figure}
\includegraphics[width=1.0\textwidth]{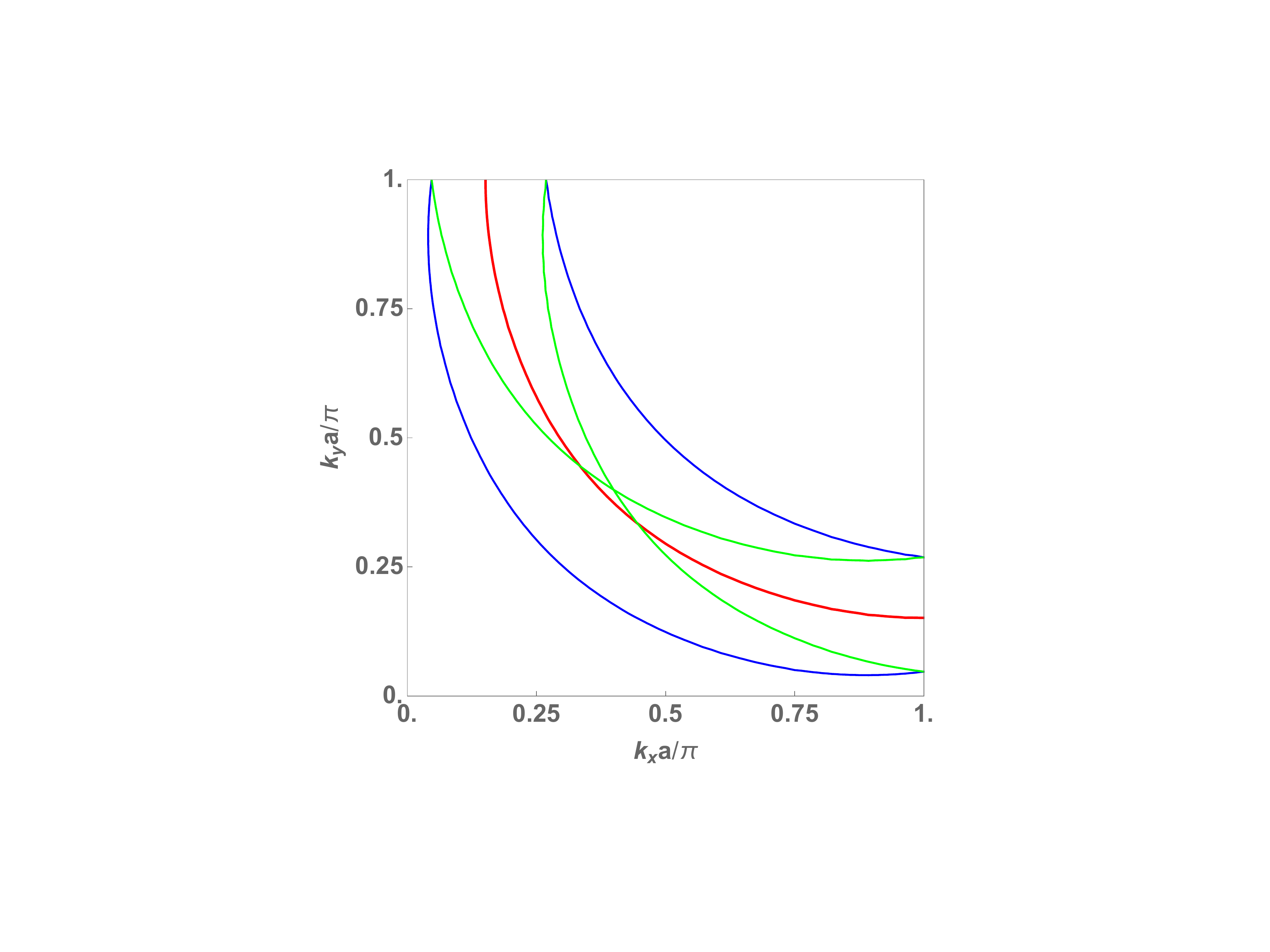}
\caption{The un-reconstructed or erstwhile Fermi-surface in a quarter of the BZ is shown in red for a hole-doping of about 12\%. The four nearest ghost Fermi-surfaces displaced by the mini-reciprocal vectors $ (\pm 2 \pi/(2Pa), \pm 2 \pi/(2Pa))$ from it and for $P=9$ are also shown. Their spectral weight in the absence of scattering is zero. The phase space for scattering between the states associated with the principal mBZ  and any of these nearby mBZs occurs with momentum transfer $(\pm 2 \pi/(2Pa), \pm 2 \pi/(2Pa))$ and is restricted to the region between them by Pauli blocking.}
\label{FSs}
\end{figure}

At $T=0$, Pauli blocking allows only scattering allowed between states $|{\bf k+g_p}>$ and $|{\bf k + g}'>$ for which either $\epsilon_0({\bf k+g_p})$ or $\epsilon_0({\bf k+g}')$ is below the chemical potential and the other is above. The one below moves further below and the one above further above the chemical potential due to the scattering. This limits the phase space for scattering to states with momenta ${\bf p} \equiv {\bf k+g}_p$ and ${\bf  p}'\equiv {\bf k+g}_p$ to those between  $\epsilon({\bf p}) = \mu$ and $\epsilon({\bf p}') = \mu$.  The contours of constant energy in the BZ  equal to the chemical potential for ${\bf g} = {\bf g}_p$ and these four neighboring ${\bf g}'s$ are shown in Fig. (\ref{FSs}). Scattering between any pair is limited to momenta lying in the area between that pair.

Let us adopt the notation of measuring ${\bf g}_p$'s from the origin for any ${\bf k}$ so that $\epsilon_0({\bf p})$ is the band-structure for $H_0$, and denote ${\bf g'}$'s nearby to such ${\bf g}_p$'s by ${\bf g}_i, i = 1,2,...$. $\epsilon_0({\bf p+g}_i)$ is then the band-structure displaced from $\epsilon_0({\bf p})$ by ${\bf g}_i$. Summarizing the above simplifications, one needs to solve only

\be
\label{H-simple}
\big(\epsilon({\bf p}) - \epsilon_0({\bf p})\big) a_{0}({\bf p}) + \sum_n \overline{M}_{0n} a_{n}({\bf p}) &=& 0,  \\   
\overline{M}_{10} a_{0}({\bf p})+ \big(\epsilon_1({\bf p}) -  \epsilon_0({\bf p +g}_1)\big) a_{1}({\bf p +g}_1) & = &0, \nonumber \\ \nonumber
\cdot ~~~~~~~ \cdot ~~~~~~~~\cdot  &=& 0,\\ \nonumber
\cdot ~~~~~~~ \cdot ~~~~~~~\cdot  &=& 0,\\  \nonumber
  \overline{M}_{n0} a_{0}({\bf p}) + \big(\epsilon_n({\bf p}) - \epsilon_0({\bf p +g}_n)\big) a_{n}({\bf p+g}_n) &= &0.
\ee
Here $\epsilon_n({\bf p})$ are the eigenvalues of  $H_0+H_b$, when $n$ mBZ's are kept in the calculation  beside the principal mBZ and the $a_i({\bf p+g_i})$'s  are the basis set.  In the numerical calculations below we keep $n = 4$ for reasons explained above. Larger $n$ produces no qualitative and only small quantitative differences. To take care of Pauli-blocking, we have introduced $\overline{M}_{0i} = {M}_{0i}({\bf p, p+g_i})$ if $(\epsilon({\bf p}) - \mu)$ is positive/negative while $\epsilon({\bf p+g_i}) - \mu)$ is negative/positive, and is $0$ otherwise. This restricts the calculation to $T=0$. Finite temperature calculations are easy to devise but no further illumination results. (We should in-principle, redetermine $\mu$ to account for any changes in particle-hole asymmetry. This is both messy and not important for the purposes of this work.)

 There is one more important approximation and simplification which is also a limitation of the procedure used to calculate. Let $\alpha_0, \alpha_i, (i =1,..,4)$, be the eigenvectors of the system (\ref{H-simple}). For $\overline{M}_{0i} \to 0$, the only eigenvalue at any momenta is $\epsilon({\bf p}) = \epsilon_0({\bf p})$ so that  $\alpha_0 = a_0$ normalized to 1. All other $a_i$ and $\alpha_i$ are null. We wish to follow this solution as $M_{0i}$ are turned on but remain small. This procedure is sensible only so long as the amplitude of $a_i, i \ne 0$ remain of the same order as $\phi$. This requires that there be no degeneracy between $\epsilon_0({\bf p +g_i})$ and $\epsilon_{\bf p}$ or at least that
  $|\epsilon_0({\bf p +g}_i) -  \epsilon_0({\bf p})|$ be not much smaller than $\overline{M}_{0i}$. Actually in the allowed region of scattering, the matrix elements are $O(t \phi/P)$, while the energy differences are typically $t/P$. For $\phi/P << 1$, the spectral weight of states that evolve from $a_i$ corresponding to ${\bf p+g}_i)$ is therefore at most only of $O(\phi)$. 
  The general spectral function is given by
  \be
 \label{spfn}
 A({\bf p}, \omega) = \sum_{({\bf g_i = 0,1,..n })} \frac{1}{\pi}\frac{\gamma}{(\omega - \epsilon({\bf p+g}_i))^2 + \gamma^2}  |\alpha({\bf p + g}_i)|^2,
 \ee
 Here $\gamma$ represents damping. If the approximation that the amplitude of the $a_i, i \ne 0$ in $\alpha_0$ are atmost of $O(\phi)$ is correct, we need only calculate the spectral function keeping only the eigenvalue $\epsilon({\bf p})$ but keep all the $a_i({\bf p+g_i})$ in $\alpha_0({\bf p})$ to get correct answers to $O(\phi)$. We will explicitly check to make sure that the amplitudes of $a_i, i \ne 0$ in $\alpha_0({\bf p})$ are no more than $O(\phi)$ to judge the validity of the approximation.  In the approximations of the present calculation the spectral weight from states which evolve from the smallest ${\bf g}_i$'s which have spectral weight no more than of $O(\phi)$ is lost. These may be important for some purposes discussed in the concluding section. Larger ${\bf g}_i$'s have weights smaller than this by at least $O(1/P)$.
 
 It should be noted that there is never any true nesting in the problem posed; only a continuous variations in energy differences and matrix elements between states as one goes around the erstwhile Fermi-surface. No problems of multiple Fermi-surfaces  therefore arise. The states $|{\bf p+g}' >$, important for scattering, track the erstwhile Fermi-surface at all angles to $O(1/2P)$, as in Fig. (\ref{FSs}). This is the next best thing to the nesting of time-reversed states as in superconductivity and produces similar effects in the single-particle excitation spectra.  There is however weight of $O(\phi)$ acquired by the nearest ghost Fermi-surfaces, which may be detectable by sensitive ARPES measurements, (see the discussion at the end).

\begin{figure}
\includegraphics[width=0.9\textwidth]{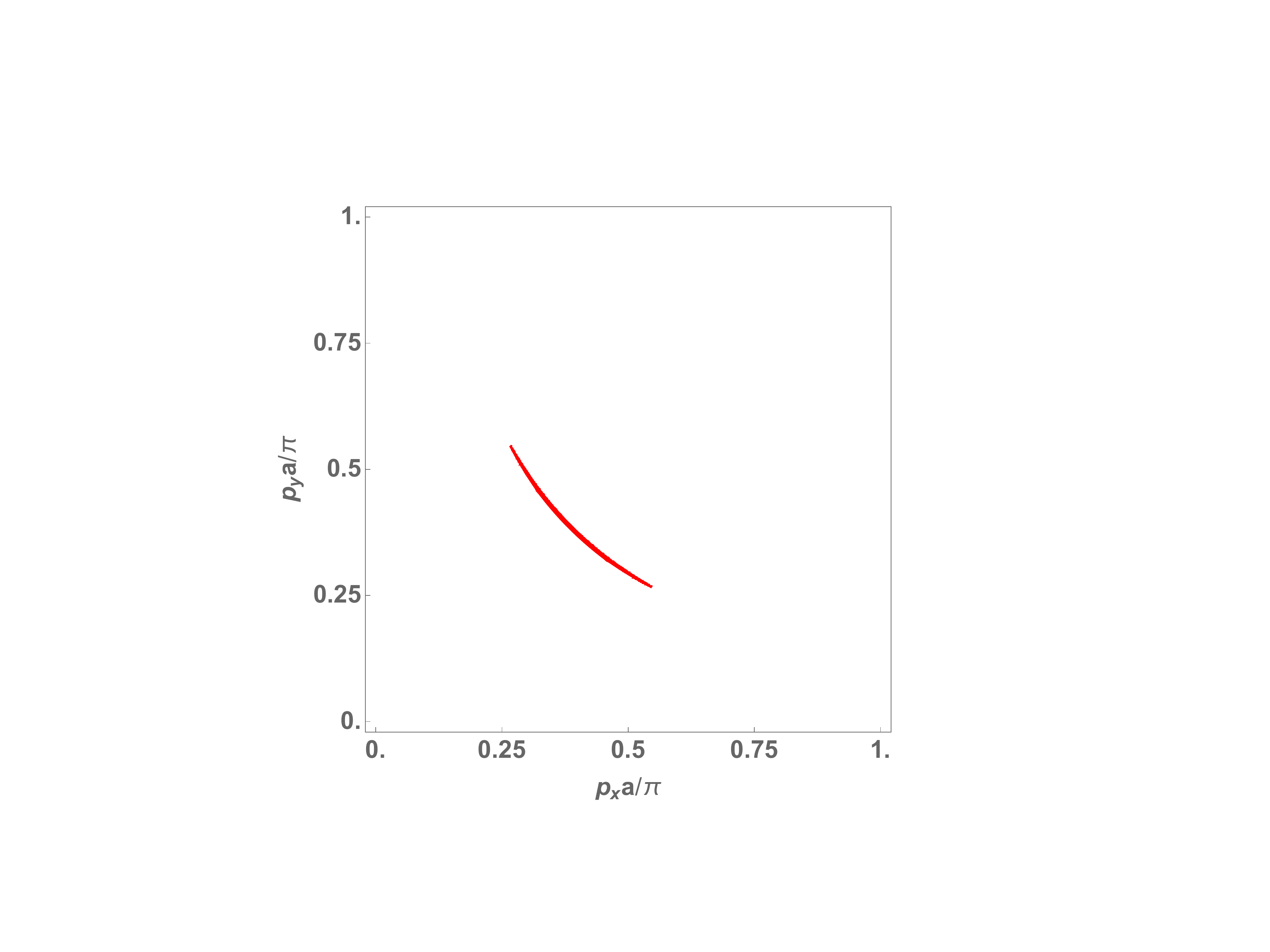}
\caption{The contour of the  spectral function (from its maximum amplitude at the point in the diagonal to about 5\% of it ) at the chemical potential in a quarter of the Brillouin zone showing the transformation of the Fermi-surface to an arc due to scattering between the principal Fermi-surface and the ghost Fermi-surfaces. A coupling constant $2\phi/P = 0.1$ and damping $\gamma = 0.03 t$ are used in this calculation. For zero damping, the arc shrinks to a point. This figure  does not show the variation of the spectral weight as a function of angle at the chemical potential. Quantitative spectral functions for the Fermi-arc and at larger angles is given in next figure (\ref{A(pf,w)-angle}).}
\label{ARC}
\end{figure}

\begin{figure}[ht]
\begin{center}
\includegraphics[width=0.9\textwidth]{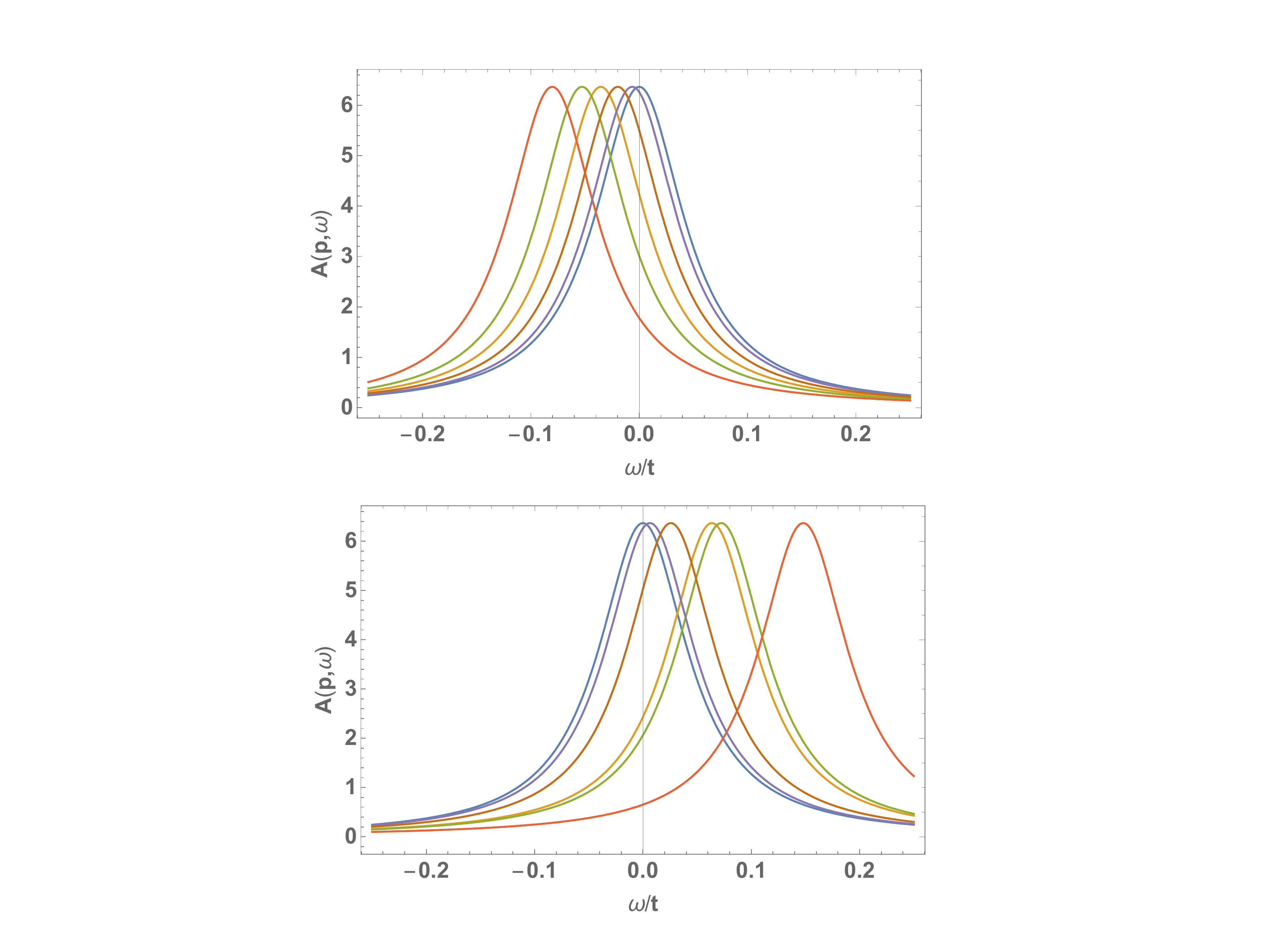}
\end{center}
\caption{{\it Top}: The spectral function at the chemical potential is shown at various angles (diagonal, and 5, 10, 15, 25, and 45 degrees from the diagonal) for a momenta infinitisimaly on the hole side of the erstwhile FS. $2\phi/P = 0.1, P=9$. A line-width $\gamma = 0.1$ is used. The diagonal is unaffected by the interactions. The arc shown in Fig. (\ref{ARC}) reflects the weight at the chemical potential due to finite damping.
{\it Bottom}:  The same as above but with momenta infinitesimally on the particle side of the erstwhile FS. Particle-hole symmetry is lost.}
\label{A(pf,w)-angle}
\end{figure}

 Numerical results for the spectral function showing the pseudogap and the Fermi-arc will be presented based on the above approximations. We may anticipate the results qualitatively by simple arguments keeping in mind  Fig. (\ref{FSs}).  The matrix elements are non-zero only for states with momentum differences ${\bf g}_p -{\bf g}'$. At such points the energy differences are of $O({\bf v}_{Fp}\cdot ({\bf g}_p -{\bf g}'))$, which is generally of $O(t/P)$. The matrix elements connecting the degenerate points at the chemical potential shown in (\ref{FSs}) are 0.  The matrix elements increase with angle from the diagonal direction  towards the $(0,\pi)$ directions, where they are $t\phi/P$. Moreover the Fermi-velocity is 3 to 4 times smaller in the $(0,\pi)$ directions than in the diagonal direction. 
 Special attention should be paid to the diagonal directions ${\bf p}_x = \pm {\bf p}_y$ near the erstwhile FS.  Two of the four states ${\bf g}_{i,x}= \pm{\bf g}_{i,y}$ have matrix elements exactly zero.  One can see by examining Fig. (\ref{FSs}) that the other two are Pauli-blocked. $\overline{M}_{0,i} =0$ for points on the FS in the diagonal direction. One can check that this is generally true irrespective of the number of ${\bf g}_i$'s kept, except for symmetry related points on the erstwhile FS which are connected with very large differences of $\nu$'s. 
From these considerations, we should expect a gap of $O(t \phi)$ in the $(0,\pi)$ directions decreasing to 0 in the diagonal direction. At an angle $\theta$ to the diagonal, the expected gap is of $O(t \phi^2 \theta^2/P$.  A Fermi-arc is expected for angles in which this is smaller than the line-width. In the pure limit,$\gamma =0$, there are only Fermi-points.

\begin{figure}[ht]
\includegraphics[width=1.0\textwidth]{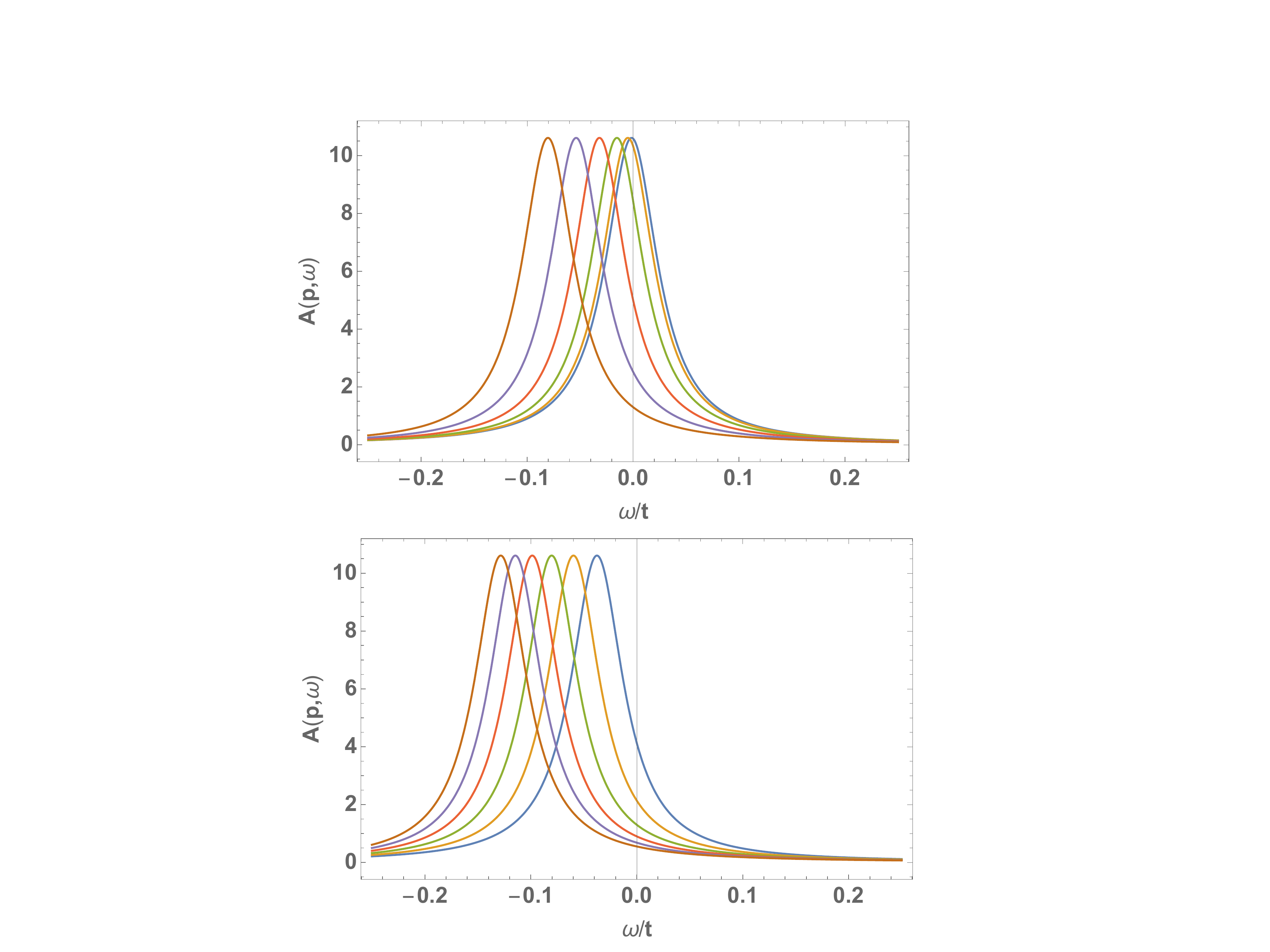}
\caption{Figure showing the effect of the coupling constant and of the period 2P of the loop-current order.
Top: The hole spectral function as a function of energy at the erstwhile FS point in the $(0,\pi)$-direction with change of coupling constant $2\phi$ from $0$ to $0.1$ in steps of $0.02$ for a fixed P=9.
Bottom: The same for varying period P from 5 to 15 in steps of 2 with fixed coupling constant $2\phi = 0.1$. The gap increases with $P$. A damping $\gamma= 0.05 t$  is used.}
\label{var-angle,coup.const}
\end{figure}
 
 \begin{figure}[ht]
\includegraphics[width=1.0\textwidth]{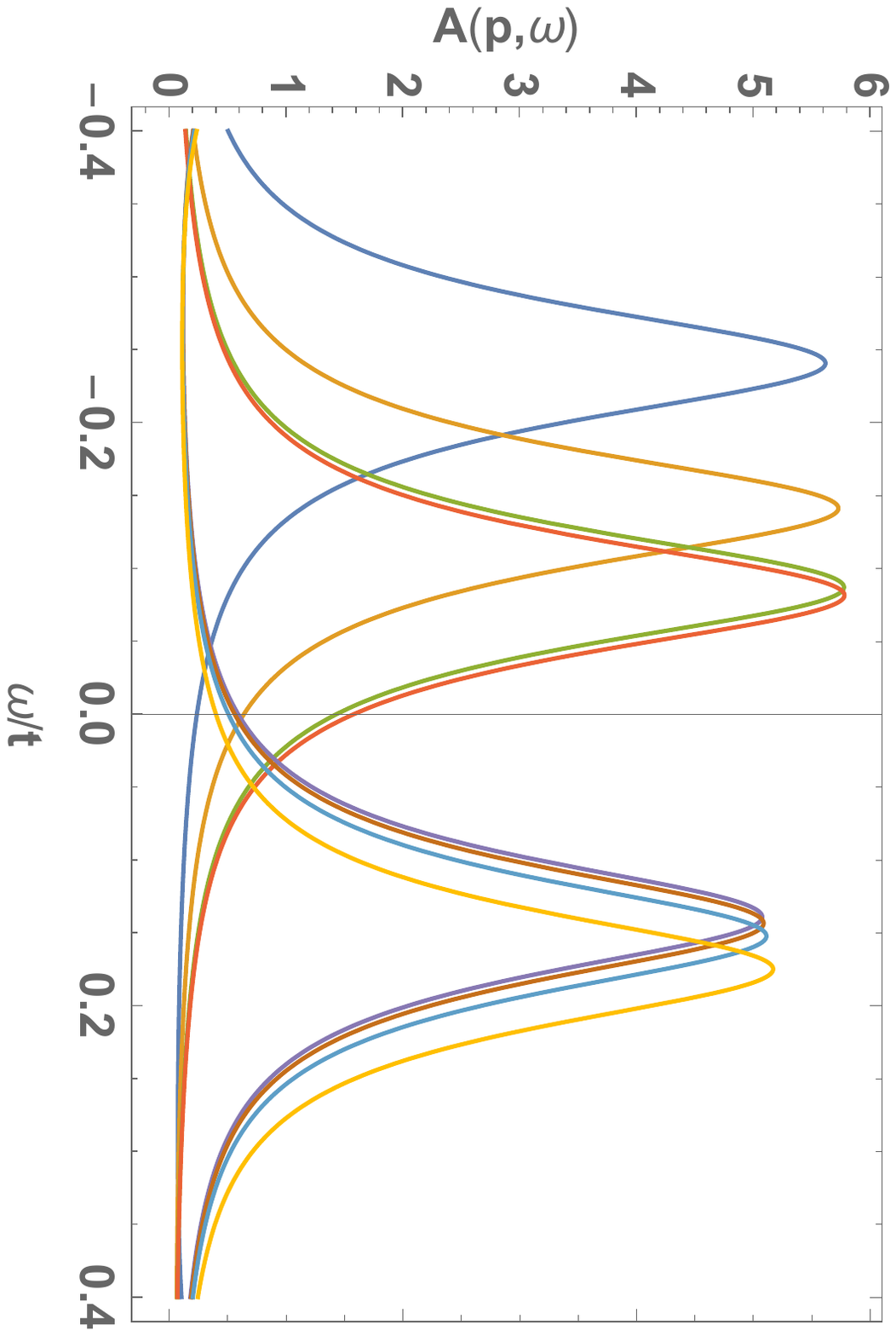}
\caption{The spectral function $A({\bf p},\omega)$ is shown as a function of $\omega$ for various ${\bf p}$ on either side of the erstwhile FS ${\bf p}_{Fx} = 0.151 \pi/a$ in the $(0, \pi)$ direction. From the left, the curves are at ${\bf p}_{x} = (0.12, 0.14, 0.150, 0.151, 0.152, 0.153, 0.155, 0.16)\pi/a$. Calculations are for $2\phi/P =0.05$ and $P=9$.}
\label{Fig(vary p)}
\end{figure}

 \subsection{Calculated Results for one particle spectral function.}

 We are interested in the spectra close to the Fermi-surface ${\bf p}_F \equiv {\bf k_F + g}_p$. As explained in the last section, one needs consider states only  for $\nu'_x - \nu_{px} = \pm 1, \nu'_y - \nu_{py} = \pm 1$. Explicit calculations with more states changes results negligibly.
 The un-reconstructed Fermi-surface ${\bf p}_F$ for cuprate at a doping of about 12\% for a quarter of the BZ is shown in Fig. (\ref{FSs}) together with the four nearest shadow Fermi-surfaces, displaced to ${\bf p}_F + {\bf g}'$, with ${\bf g}'_{x,y} - {\bf g_p}_{x,y} = \pm 2 \pi/(2Pa)$. There is no scattering between the three degenerate crossing points near the diagonal (barring effects due to disorder), because  these points are in different mBZS such that ${\bf k'+g}'= {\bf k"+g}"$. This implies ${\bf k'} \ne {\bf k"}$ which is not allowed. These degeneracies are therefor not lifted. For the same reason, the amplitudes of states of the ``ghost" fermi-surfaces at the diagonal remain 0 and that of the principal Fermi-surface remains 1. 
  
   \begin{figure}
\includegraphics[width=1.0\textwidth]{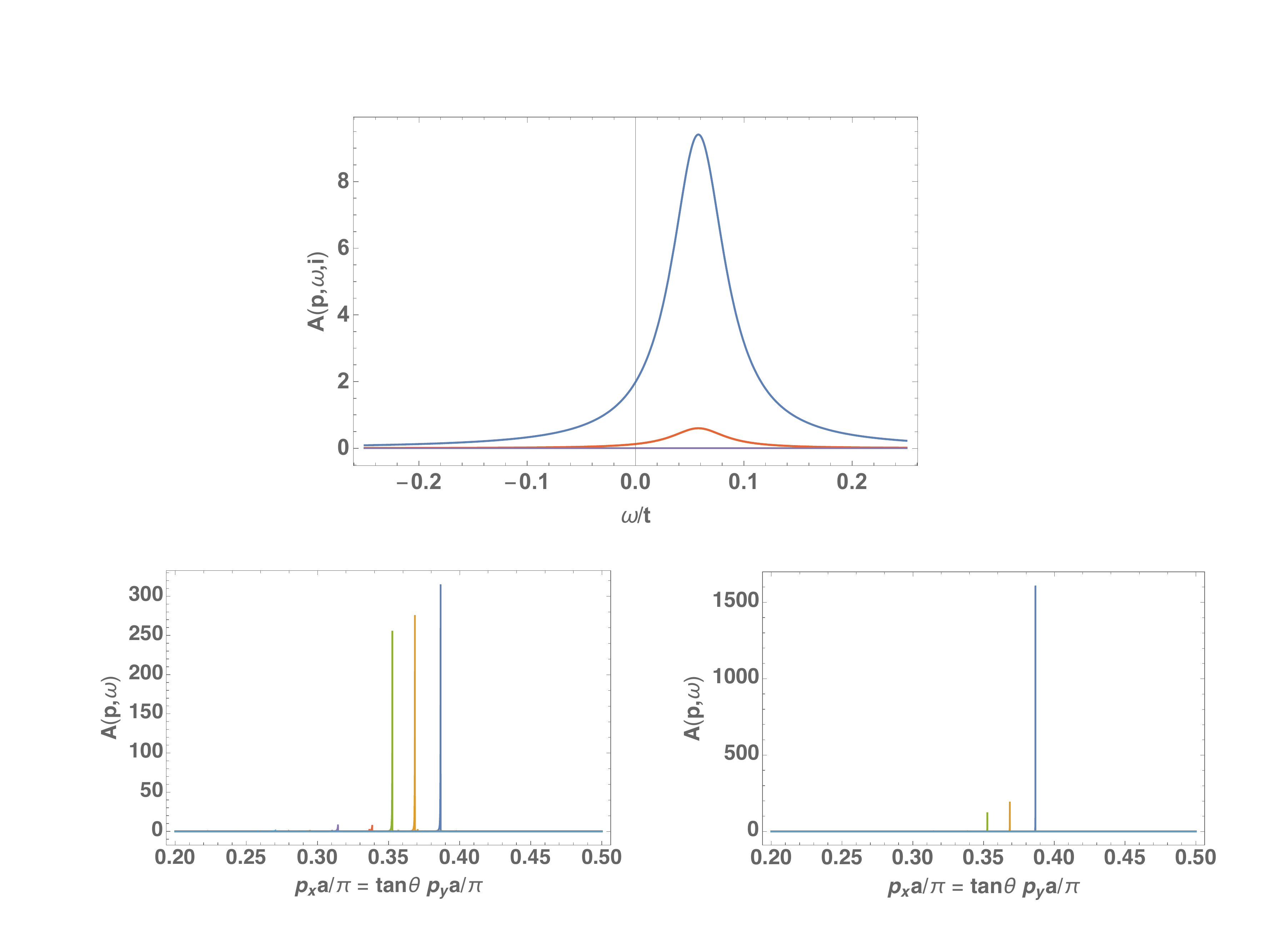}
\caption{Top: The partial spectral weight at the anti-nodal Fermi-point is shown as a function of energy for the principal mBZ and for the four nearby mBZs (two of which have zero spectral weight and the other two have identical spectral weight by symmetry of scattering). The coupling constant $2\phi/P =0.05$ and $P=9$. The principal mBZ has spectral weight more than 0.9 of the total; by varying $\phi$, it can be verified that this weight is less than 1 by $O(\phi/P)$. These results are important to show the validity of the approximations in this work.\\
Bottom: The momentum distribution curve for $2\phi/P=0.05, P=9$ is shown on the Fermi-surface $\omega = \mu$ at the diagonal of the BZ and at angles separated by $\arctan(1/1.n)$, n=1,2,... from it at damping $\gamma = 0.005$ on the left and 
$\gamma = 0.0001$ on the right. These angles correspond approximately to 2 degree successive departure from the diagonal. This may be compared to Fig (\ref{ARC}) in which $\gamma = 0.03$ and the Fermi-arc has comparable weight over nearly 30 degrees.}
\label{Fig-Tests}
\end{figure} 

In Fig.(\ref{ARC}), the contour of the spectral function at the chemical potential from the calculation of Eq. (\ref{spfn}) is shown for a finite damping.  The promised Fermi-arc is obtained. The angle of the Fermi-arc decreases as the coupling constant $\phi$ is increased and increases as $2P$ or the damping is increased. Since the contour does not represent the distribution of the spectral weight along it quantitatively, the spectral function is explicitly shown in Fig. (\ref{A(pf,w)-angle}).  In the diagonal directions, the spectral function does not change due to the scattering. At larger angles it develops a gap which grows continuously towards the $(0,\pi)$ directions. For any given damping, the spectral weight at the chemical potential for ${\bf p}_F$ decreases at larger angles displaying the Fermi-arc shown in Fig(\ref{ARC}).  The gap as a function of angle rises at small angles somewhat more slowly than the BCS d-wave energy function primarily due to the damping introduced.  In the limit of zero damping only a Fermi-point is obtained as is shown below.
 
In Fig. (\ref{var-angle,coup.const}) the spectral function is plotted for various $\phi/P$ for fixed $P$ and various $P$ for fixed $\phi/P$ at the erstwhile Fermi-surface in the $(0,\pi)$ direction. The increase of  the gap at the chemical potential in the $(0,\pi)$ as the coupling constant is increased may be taken to simulate the increase of the gap as the temperature is decreased below $T^*(x)$. The calculations are for $T=0$. Independently of the coupling constant increasing temperature would tend to decrease the gap due to the usual Fermi-factors. 

In Fig. (\ref{Fig(vary p)}) the spectral function is plotted for various ${\bf p}$ on either side of ${\bf p}_F$ in the $(0,\pi)$ direction. Far enough away from ${\bf p}_F$, the spectral function is unaffected by scattering as the phase space for particle-hole scattering among the principal and shadow FSs vanishes. This is controlled by the period $2P$ being larger for smaller $P$. A particle-hole anisotropy is also found in the spectral function related to the difference of phase space for particle and for hole scattering \cite{Johnson2010}.

\begin{figure}[ht]
\includegraphics[width=1.0\textwidth]{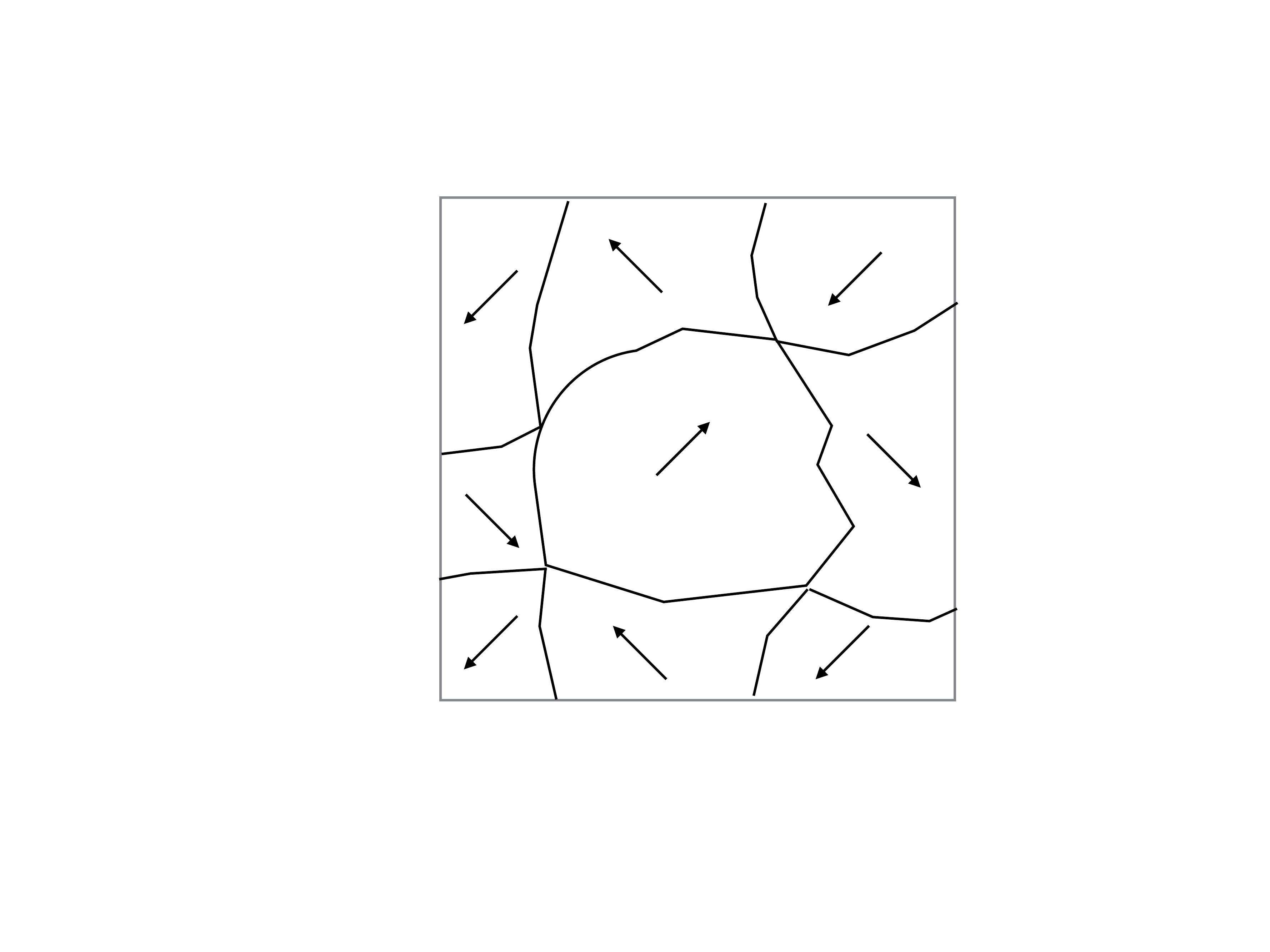}
\caption{Adiabatic deformation of the periodic domains of Fig. (\ref{Fig2}) by weak disorder preserving the essential topology of the domain boundaries. This is a variant of the four-color problem in which two pairs  of  four colors  meet only at points while the other two have non-zero boundaries. Then, it appears that filling of any arbitrarily large area with these rules enforces vortex structure at the points of intersection of the domains.}
\label{Domains-disord}
\end{figure}

A test of the approximations used is shown in the top in Fig. (\ref{Fig-Tests}). The partial spectral weight in the eigenvector of the principal mBZ and the four nearby mBZs is shown at the anti-nodal point. The latter is systematically
$O(\phi)$ of the total spectral weight. At the bottom of the same figure, the spectral function at the erstwhile Fermi-momentum at different angles is shown at $\omega = \mu$ and for two different small damping coefficients $\gamma$.
The spectral weight is increasingly concentrated along the diagonal direction as the damping is decreased so that in the pure limit only a Fermi-point remains.

All essential features of the observed one-particle spectra and of the Fermi-arcs appears to be reproduced by the calculations presented above. These include the angle dependence of the gap, the shrinking of the arcs as one moves below $T^*(x)$ to their saturation to a finite angle, and the diminishing weight at the chemical potential of the momentum distribution curve as temperature decreases \cite{Kaminski-pg}. One feature to which attention has not been paid here is the apparent spatial variation in the magnitude of the gap and the increased linewidth of the spectra in the pseudogap state. These are obviously features associated with the increased coupling to disorder. It is tempting to associate them with the fact that disorder couples linearly to the inversion odd part of the loop-current order parameter favoring one over another domain. Such a coupling would make the domains vary in size but it is expected that the essential topological features of the domain boundaries, evident in Fig. (\ref{Fig2}), would be preserved, as is sketched in Fig. (\ref{Domains-disord}). This would in general give rise to spatial variation in the size of the pseudogap and associated increase in the elastic scattering rate. It is worthwhile noting that the information on variation of pseudogap etc., which comes from STM measurements \cite{Hoffman2002}, \cite{Davis} comes mostly from experiments in Bi2212, which is more disordered than cuprates such as YBCO. An interesting aspect of disorder is that it would relax the momentum conservation rules in scattering so that crossings  such as in Fig. (\ref{FSs}) could change to closed surfaces. Occasionally Fermi-arcs have been seen touching small broadened Fermi-surfaces with much smaller spectral weight on one side \cite{XJZhouSmallFS}. This is especially interesting in relation to the sketch made in Fig. (\ref{FS(H)}). 
Related observations include Ref. \cite{Johnson2010} and ARPES in at least one electron-doped cuprate beyond its AFM region \cite{Shen-edoped}. 

\subsubsection{Estimate of the  Period}
The period $2P$ and the phase $\phi$ should in principle be be calculated variationally. 
 Although it was argued that the spectral function can be calculated correctly to $O(\phi)$ in the approximations used, one cannot calculate the change in energy correctly to the same order because the gaps are at most of $O(\phi)$ so that the change in energy if of $O(\phi^2)$. So the diagonalization of a massive matrix is required to variationally determine $P$ and $\phi$. One may make a rough guess at the period $P$ using the value for the phase $2\phi$ consistent with the moment which has been deduced in polarized neutron scattering measurements and the assumption that it does not change significantly if $P$ is large enough. A staggered magnetic moment of 0.1 to 0.2 $\mu_B$ in a  unit-cell corresponds to $2\phi/P \approx$ 0.05 to 0.1. From the calculations above, the observed gap at the "anti-nodal point" of about 0.1$t$ requires $P$ of between $6$ and $15$.
 
\subsubsection{Earlier theories for the pseudogap} 

Calculations for the pseudogap have mostly followed the idea of antiferromagnetism or antiferromagnetic fluctuations \cite{Tremblay} or mathematically related idea of d-wave density order \cite{Chakra-ddw} or fluctuations of gauge fields by themselves \cite{Lee-Nagaosa-rev} or coupled to antiferromagnetism \cite{Georges-Sachdev}.
They all have the problem of several Fermi-surfaces or no Fermi-arc or both. Very impressive calculations yielding results which are doubtless correct for the Hubbard model have been done \cite{Georges-pseudogap}  and yielded peaks in the self-energy in the $(0,\pi)$ directions which lead to a large reduction of the density of states in that direction. But they cannot at present be followed to around the Fermi-surface. If the large peak in self-energy is due to significant anti-ferromagnetic correlations \cite{Tremblay}, one expects other Fermi-surfaces and not a Fermi-arc or a Fermi-point.
An earlier idea proposed by me based on disordered domains of loop-current order \cite{CMV-Domains2014} taking into account almost singular forward scattering at the domain walls does have an angle-dependent peak in the self-energy  but there is no reason for  Fermi-arcs or points. As shown above, domains with specific boundary currents such as proposed here is consistent with experiments.

\section{Symmetry of the proposed order and Experiments}

In this section, the symmetry classification of the order and its change from the previously proposed symmetries is discussed. This is to re-examine the experiments already found to be consistent with loop-current order and to ask what experiments may rule out or verify the modifications proposed here.

The symmetry group of the four different orientations of $\Omega$ in the super-cell is different from that of the boundary currents.
The point group symmetry is lower than that allowed in the original loop order state ($\underline{m}mm$ which had two fold rotations about the y axis $C_y$, reflections in the xy and yz planes ( $\sigma_z$ and $\sigma_x$), inversion followed by time-reversal $RI$, reflection in the xz plane followed by time reversal ($\sigma_yR$), two fold rotation about the x-axis followed by R ($C_yR$) and twofold rotation about z axis followed by R ($C_{2z} R$). 
The only symmetry element that are left now are identity, four fold rotation about the c-axis followed by time-reversal $C_{4z}R$ and two fold rotation about the c-axis (or what is the same thing inversion about the center of the unit-cell.)  Time reversal combined with the two fold rotation or inversion is not a symmetry. No reflections on the a, b or c axis or $a\pm b$ axis are symmetries. The four-fold rotations is of-course only allowed in a tetragonal crystal and not in an orthorhombic or mono-clinic crystal. From Birss classification \cite{Birss} (Table 3 - page 330), this belongs to the group $\underline{4}$ for a tetragonal crystal and $m$ for the orthorhombic or mono-clinic crystal. $m$ has elements $1, \bar{2}$. (bar over 2 because it is assumed that there is a reflection plane normal to the c-axis.) If this is missing, it would be simply have the elements $1,2$ and would be called $2$. 

The group $m$ has independent non-zero magneto-electric tensor elements: $Q_{13},Q_{23},
Q_{31}, Q_{32}$. These are the same elements as in $\underline{2}/m$ favored in (multi-domain) second harmonic generation in a mono-clinic crystal \cite{Hsieh2017}. Antisymmetric component between the elements is allowed. The observed Polarization rotation with simultaneous rotation of the principal axis for light propagation is consistent with the new symmetry also, as it was with the earlier symmetry.. These are the same elements as in $\underline{2}/m$ favored in (multi-domain) simple harmonic generation. 

Neutron scattering has already been discussed in the previous section. It is straightforward to see that dichroic ARPES \cite{Kaminski-diARPES} also arises in the proposed new symmetry, since it requires only an anti-symmetric part of the magneto-electric tensor.

The necessary non-zero elements of the third order electric dipole susceptibility $\chi_{ijk}$ which are a requisite for the observed second harmonic generation are non-zero in $m$ also. However, there were aspects of the second harmonic generation which were not adequately explicable \cite{Hsieh2017} in the previous analysis based on $\underline{2}/m$. It will be interesting to check if the lowered symmetry and the extra independent elements in $\chi_{ijk}$ provided by $m$ fit the details better. 

Local experiments such as $muSR$ \cite{muons1210} which do see time-reversal breaking signatures but with a slow fluctuation rate $10^{-5}$ slower than the pseudogap time-scale (and not infinitely slow as for truly static order) would present the same confirmation and pose the same problem. It would be interesting to investigate if the slow time-scale can arise from low frequency quantum-fluctuations of the domain walls in the proposed super-cell. 

The angular variation of the magnetic susceptibility \cite{Matsuda-torque1, Matsuda-torque2} discovered below $T^*(x)$, the ultrasound attenuation singularity at $T^*(x)$, the change in magnetic susceptibility and transport starting at $T^*(x)$ are consistent with the proposed new order as well as they were with the loop-current order proposed earlier. 


Any experiments which are related to the quantum-critical fluctuations of the loop-current order are expected to essentially remain unchanged by the variations introduced by the new order except in the long wave-length limit. The introduction of a new large period inside of which the loop-current order exist at the unit-cell level is expected to change the fluctuations only to $O(1/(2P)^2)$ in momentum space. The experiments in ARPES which directly observe the single-particle marginal Fermi-liquid relaxation rate proportional to $\omega$ and nearly angle independent, derived from the fluctuations integrated over all momenta are expected to be unaffected at least to this order and so are therefore the predicted $T \log T$ specific heat \cite{Taill-spht} and the linear in $T$ resistivity and other properties which follows from them. There are other aspects of experiments such as the large thermal Hall effect \cite{Taill-thhall} which need investigation in view of the large loops of currents in the proposed order.

{\it Magneto-oscillations}
\begin{figure}[ht]
\includegraphics[width=1.0\textwidth]{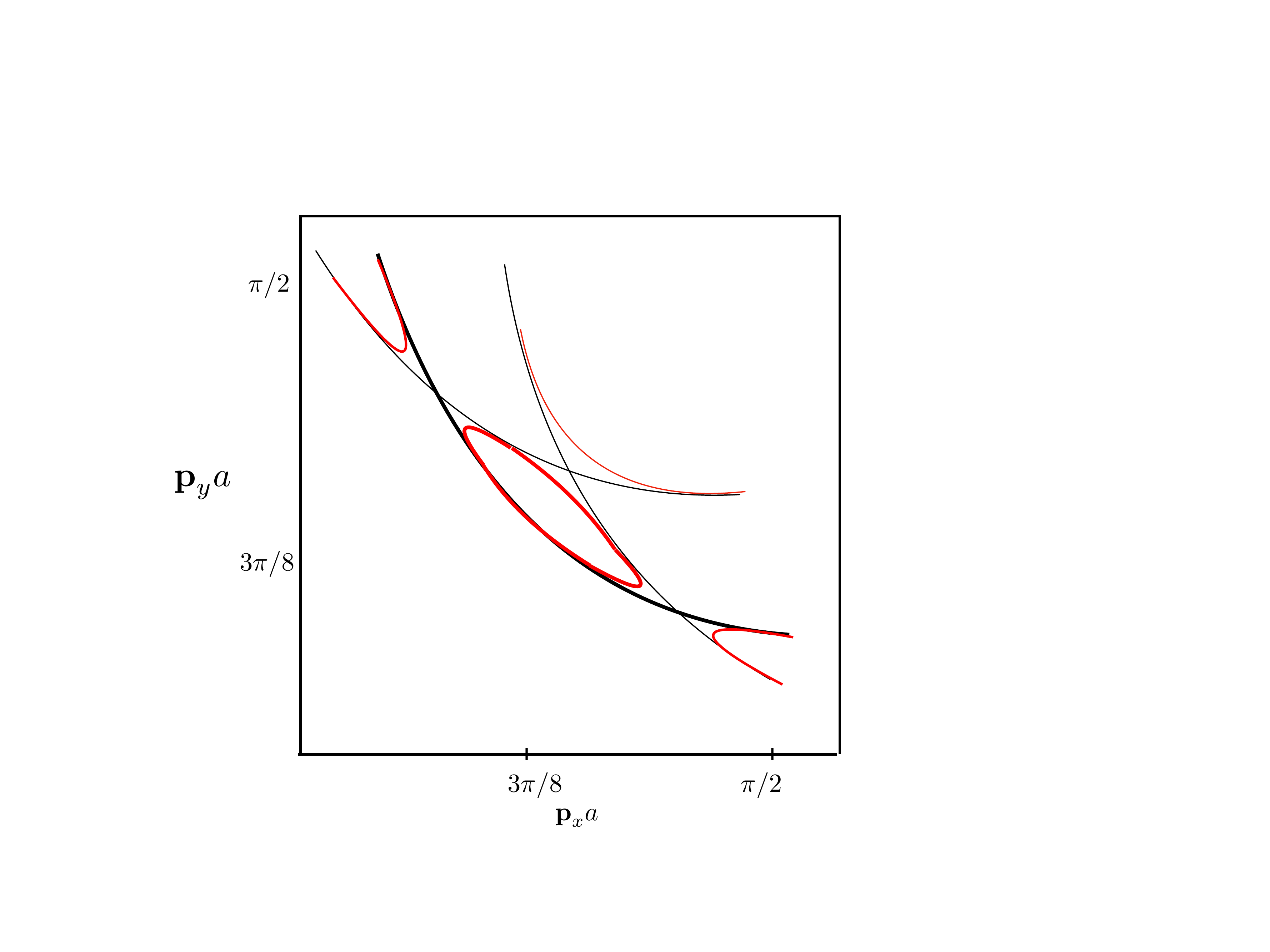}
\caption{A sketch (red) of the possible reconstruction of the arc of the principal Fermi-surface and the arcs of the nearby shadow Fermi-surface of much smaller spectral weight into a continuous curve in a magnetic field. P =9 and $2 \phi/P=0.05$ is used in the calculations.}
\label{FS(H)}
\end{figure}

The best current idea for magneto-oscillatory phenomena \cite{Proust-Sebastian2015} requires a  Fermi-arc and the pseudogap, and then connects up the arcs through a charge density wave (CDW) which introduces a new period. The general idea appears quite sound although it has not been entirely clear that the observed CDW have the requisite $Q's$ to accomplish the purpose. If that is the applicable theory, the present theory provides the necessary ingredients of  the Fermi-arc and the pseudogap.

An interesting possibility to investigate in this regard is what happens to the Fermi-arc and the shadow bands which intersect when a magnetic field is applied even without a CDW. A magnetic field does not conserve  momentum  but conserves energy. The degeneracy at the intersections between the principal Fermi-surface and the ghost Fermi-surfaces at the chemical potential must then be lifted in a magnetic field. What form the resulting shapes at the chemical potential take is an interesting question worth investigation. In the calculations above only the reconstruction of the principal Fermi-surface to Fermi-arcs has been explicitly shown. It also follows that the nearby ghost Fermi-surfaces also turn into arcs, with much reduced spectral weight).  It is interesting to note that the intersection of the Fermi-arc with the arcs as observed and of the nearby ghost Fermi-surfaces  encloses an area which is similar in magnitude to that of the small Fermi-surface deduced in the magneto-oscillation experiments for parameters giving the right magnitude of pseudogap and the angle of the Fermi-arc. A speculative sketch of such a reconstruction is shown in Fig. (\ref{FS(H)}). However, the arcs are finite only due to elastic or inelastic scattering and how this plays with the magnetic field and yet give observable oscillations must be worked out. Without a magnetic field applied, some evidence has been presented  \cite{XJZhouSmallFS} bearing resemblance to Fig. (\ref{FS(H)}) suggesting that the principal feature, the arc, is attached to a small elliptical Fermi-surface with much smaller quasi-particle amplitude.

These and the other issues in relation to experiments are worth detailed examination only if the requisite sufficiently high resolution diffraction or imaging experiments on high quality samples reveal a new periodicity. If so, each of the three principal new phenomena discovered in the cuprates may have found a satisfactory theory.

{\it Acknowledgements}: I have benefitted from discussions with Dung-Hai Lee and Lokman Tsui on the ideas presented here, and with Philippe Bourge, Adam Kaminski, Steve Kivelson, Yvan Sidis, Louis Taillefer, Suchitra Sebastian,  Z.-X. Shen, and Xingjiang Zhou on discussions of experiments. Instructions by Lokman Tsui and by Debanjan Chowdhury on numerical methods were essential. I also wish to thank the Aspen Center for Physics and the staff and faculty of condensed matter at NYU and at Berkeley where this work was done, for their courtesy.

\newpage

\bibliographystyle{apsrev4-1}

\end{document}